\documentclass[journal]{IEEEtran}
\newcommand{\ignore}[2]{ \hspace{0in}#2}

\usepackage[centertags]{amsmath}
\usepackage[figuresright]{rotating}
\usepackage{parskip}
\usepackage{calc}
\usepackage{amsfonts}
\usepackage{booktabs}
\usepackage{supertabular}
\usepackage{tabularx}
\usepackage{multirow}
\usepackage[export]{adjustbox}
\usepackage{array}
\usepackage{booktabs}
\usepackage[latin1]{inputenc}
\newcolumntype{C}[1]{>{\centering\arraybackslash}p{#1}}
\newcolumntype{M}[1]{>{\centering\arraybackslash}m{#1}}

\usepackage[]{siunitx}
\usepackage{graphicx}
\usepackage{graphbox}
\usepackage{comment}
\usepackage[hyphens]{url}
\usepackage[plainpages=false,pdfpagelabels=true]{hyperref}
\hypersetup{
    colorlinks,
    citecolor=black,
    filecolor=black,
    linkcolor=black,
    urlcolor=black,
	pdfprintscaling=None
}
\usepackage{pdfcomment}

\renewcommand{\marginnote}[2]{ \hspace{0in}#2} 

	\usepackage[backend=biber,
    style=ieee,
    sortlocale=en_US,
    natbib=true,
	maxbibnames=100,
    maxcitenames=2,     
    mincitenames=1,     
    url=true,           
    doi=false,
	isbn=false,       
 	firstinits=true, 
    eprint=false,
	uniquename=mininit,
    uniquelist=false     
]{biblatex}

   \usepackage{csquotes}

   \DeclareNameAlias{sortname}{last-first}
   \DeclareNameAlias{default}{last-first} 
	\AtEveryBibitem{\clearfield{month}}
\AtEveryBibitem{\clearfield{day}}
   \DeclareFieldFormat*{url}{}            
   \DeclareFieldFormat*{language}{}       
   \DeclareFieldFormat*{note}{}           
   \DeclareFieldFormat[misc]{url}{\mkbibacro{URL}\addcolon\space\url{#1}}
	 \DeclareFieldFormat[misc]{booktitle}{\addcolon\space #1} 
	 \DeclareFieldFormat[book]{edition}{#1 edition} 
   \DeclareListFormat[inproceedings]{address}{}  
   \DeclareListFormat[inproceedings]{location}{} 
	 \DeclareListFormat[inproceedings]{publisher}{}
   \DeclareListFormat{language}{}           
   \DeclareFieldFormat*{urldate}{}
   \DeclareFieldFormat[misc]{urldate}{\mkbibparens{\bibstring{urlseen}\space#1}}
	 \DeclareFieldFormat[article]{volume}{\bibstring{volume}\addnbspace #1}
   \DeclareFieldFormat[article]{number}{\bibstring{number}\addnbspace #1}
   \renewbibmacro*{volume+number+eid}{
      \setunit{\addcomma\space}
			\printfield{volume}
      \setunit{\addcomma\space}
      \printfield{number}
      \setunit{\addcomma\space}
      \printfield{eid}}
	\usepackage{xpatch}
	
	\DeclareSourcemap{
		\maps[datatype=bibtex]{
		\map{
			\step[fieldsource=url, match=\regexp{\{\\_\}}, replace=\regexp{_}]
		}
	 }
	}

	\AtEveryBibitem{
    \ifentrytype{inproceedings}{
			\clearlist{publisher}
		}{
     }
  }
	
	\xpatchbibdriver{misc}
  {\usebibmacro{title}}
  {\usebibmacro{title}
   \newunit\newblock
   \usebibmacro{booktitle}}
  {}{}

\begin{document}
\title{Towards a Functional System Architecture for Automated Vehicles}
\author{Simon~Ulbrich,
        Andreas~Reschka,
				Jens Rieken,
				Susanne Ernst,
				Gerrit Bagschik,
				Frank Dierkes,
				Marcus Nolte,
        and Markus~Maurer
\thanks{S. Ulbrich was with the Institute of Control Engineering at TU Braunschweig, Germany and is now with Audi AG.}
\thanks{A. Reschka, J. Rieken, S. Ernst, G. Bagschik, F. Dierkes, and M. Nolte are with the Institute of Control Engineering at TU Braunschweig, Germany. A. Reschka is currently visiting student researcher at Stanford University.}
\thanks{M. Maurer is professor and head of the Institute of Control Engineering at TU Braunschweig, Germany.}
\thanks{Manuscript received March 10th, 2017.}}

\maketitle

\begin{abstract}
This paper presents a functional system architecture for an automated vehicle. 
It provides an overall, generic structure that is independent of a specific implementation of a particular vehicle project. 
Yet, it has been inspired and cross-checked with a real world automated driving implementation in the Stadtpilot project at the Technische Universit\"at Braunschweig. The architecture entails aspects like environment and self perception, planning and control, localization, map provision, Vehicle-To-X communication, and interaction with human operators. 
\end{abstract}


\IEEEpeerreviewmaketitle

\section{Introduction}
\IEEEPARstart{W}{hen} software is developed for an automated vehicle, it is a bottom-up process for many teams. If existing building blocks are just \textit{hacked} together in some way, this will lead to complex system designs. 
Yet, having a well structured functional system architecture is key. It has central impact on the system design and technical software development for an automated vehicle --- often for several years.

This paper presents an overall functional system architecture for an automated vehicle. Implementation independent modules are grouped such that there are clean interfaces among these modules.

This functional system architecture differs from others by strictly using hierarchy and functional separation. 
It underwent several iterations. Earlier versions of this architecture have been published by \cite{Wille2010,Saust2010, Reschka2011,Nothdurft2011a, Wille2012,Ulbrich2014b,Rieken2015, Matthaei2015b,Matthaei2015a1}. It has strongly been influenced by the functional system architecture developed by Dickmanns' group \cite{Hock1992, Dickmanns1994,Dickmanns2007,Maurer2000}.
\citet{Matthaei2015b}, and \citet[p. 37\,ff.]{Matthaei2015a1}. 
The following architecture discussions will be based on the last architecture revision in English \citep{Matthaei2015b}.

Some refinements have already been published in German \citep[p. 37\,ff.]{Matthaei2015a1}. A goal of this article is to make our research insights accessible to the international scientific community. Apart from that, several other enhancements have been made to incorporate more recent research such as the definition of interfaces, e.g., the definition of a \textit{scene} and a \textit{situation} in \citet{Ulbrich2015d}, or functional safety considerations resulting from the application of the architecture in the aFAS project \citep{Stolte2015}.

This paper is organized as follows. First different approaches for structuring driving tasks and their processing levels are introduced and compared with other approaches in the literature. Then the functional system architecture from this paper is presented by outlining its main columns and clarifying its interfaces and comprised activities. For each aspect the modifications to the state of the art are presented and provided with a root cause for these. In the end, open issues are highlighted and a conclusion is drawn. 

\section{Background}
\label{sec:Background}
The\marginnote{Requirements} ISO 26262 standard proposes a functional system architecture as a part of the system design and defines \enquote{modularity}, \enquote{adequate level of granularity}, and \enquote{simplicity} as requirements to the architectural design \citep[part 4, p. 13]{ISO26262}. As a property of a modular system design the ISO 26262 proposes \enquote{hierarchical design}, \enquote{precisely defined interfaces}, \enquote{maintainability}, \enquote{testability}, and \enquote{avoidance of unnecessary complexity} (ibid.). 

To provide a structure for human behavior,\marginnote{Rasmussen} \citet{Rasmussen1983} distinguishes \enquote{skill-based behavior}, \enquote{rule-based behavior}, and \enquote{knowledge-based behavior} as three levels of performance of skilled human operators. On the lowest, skill-based level, reactive, sensory-motor activities take place without conscious control. On a rule-based level, decisions are taken based on a previously stored set of rules. If a situation is not familiar and there is no stored rule for it, knowledge-based behavior may be applied. Here a new strategy for goal archival is developed from existing knowledge. 

Donges\marginnote{Donges} distinguishes \enquote{navigation}, \enquote{guidance}, and \enquote{stabilization} as three hierarchical levels of driving tasks in his publication from 1982\ignore{\citet{Donges1982}} cited in \cite{Donges1999}. Similar to \citet{Riemersma1979} and \citet[p. 489]{Michon1985} citing his inaugural lecture from 1971,\ignore{\citedin[p. 4]{Muller2006}} one elementary (operational) layer is used for course keeping and speed control, a second (tactical) layer is for any behavior planning, and a third one is for strategic planning. 

\begin{table}[htbp]
	\centering
	\caption{Examples for driving tasks and their processing levels based on \citet[p. 1383]{Hale1990} depicted as in \citet[p. 8]{Muigg2009}}
	\label{tab:RasmussenDonges}
	 \vspace{-2mm}
	  \begin{tabular}{@{} M{0.1cm} M{1.4cm} | M{1.7cm}  M{1.7cm} M{1.7cm}}
		 & & \multicolumn{3}{c}{\textbf{Processing level}}\\[0.1ex]
		 & & \multicolumn{1}{c}{\textbf{Skill-based}} & \multicolumn{1}{c}{\textbf{Rule-based}} & \multicolumn{1}{c}{\textbf{Knowledge-based}} \\ \hline 
		\multirow{3}{*}{\rotatebox[origin=l]{90}{\parbox[c]{2cm}{\centering \textbf{Driving task}}}}
    &\textbf{Navigation} & Daily commute & Choice between familiar routes & Navigating in foreign town\\[5pt]
		&\textbf{Guidance} & Negotiating familiar junctions & Passing other car & Controlling a skid on icy roads\\[5pt]
		&\textbf{Stabilization} & Road following around corners & Driving an unfamiliar car & Learner on first lesson\\[5pt]
		\end{tabular}
\end{table}    

\citet[p. 1383]{Hale1990}\marginnote{Driving Tasks and Processing Levels} suggest that the three levels of driving tasks and the three levels of Rasmussen are rather orthogonal to each other. While Donges' driving tasks address \textit{what} task is to be solved, Rasmussen addresses \textit{how} it is solved.
Table~\ref{tab:RasmussenDonges} illustrates the relationship between both.
Most of the time stabilization tasks are handled on a skill-based level.
However, if for example a car is unknown to the driver, he or she may not have these subconscious skills to address the task.
Yet, there are learned rules that may be used.
If even these rules need to be formed from knowledge, because it is the first time a fresh learner drives a car, it would be knowledge-based behavior. 
On a tactical level, passing another car typically involves situation assessment and a certain amount of stored rules and experience, e.g., how much of a gap is necessary to overtake. Humans address this typically by rule-based behavior. 
Last of all, navigation may display the widest variety of processing levels in everyday driving. Blindly commuting the same street every day without looking for, e.g., changed traffic signs, may almost be skill-based behavior. It becomes rule-based, if it involves active tactical decisions between several route options, and turns into knowledge-based behavior, if a driver navigates in an unknown city for the first time. 

\newlength{\tempArchitectureCaptionWidth}
\settowidth{\tempArchitectureCaptionWidth}{Functional system architecture of an automated vehicle (beh. plan. = behavior planning, dyn. env. = dynamic environment,}
\begin{figure*}[htbp]
      \centering
			\includegraphics[width=\textwidth]{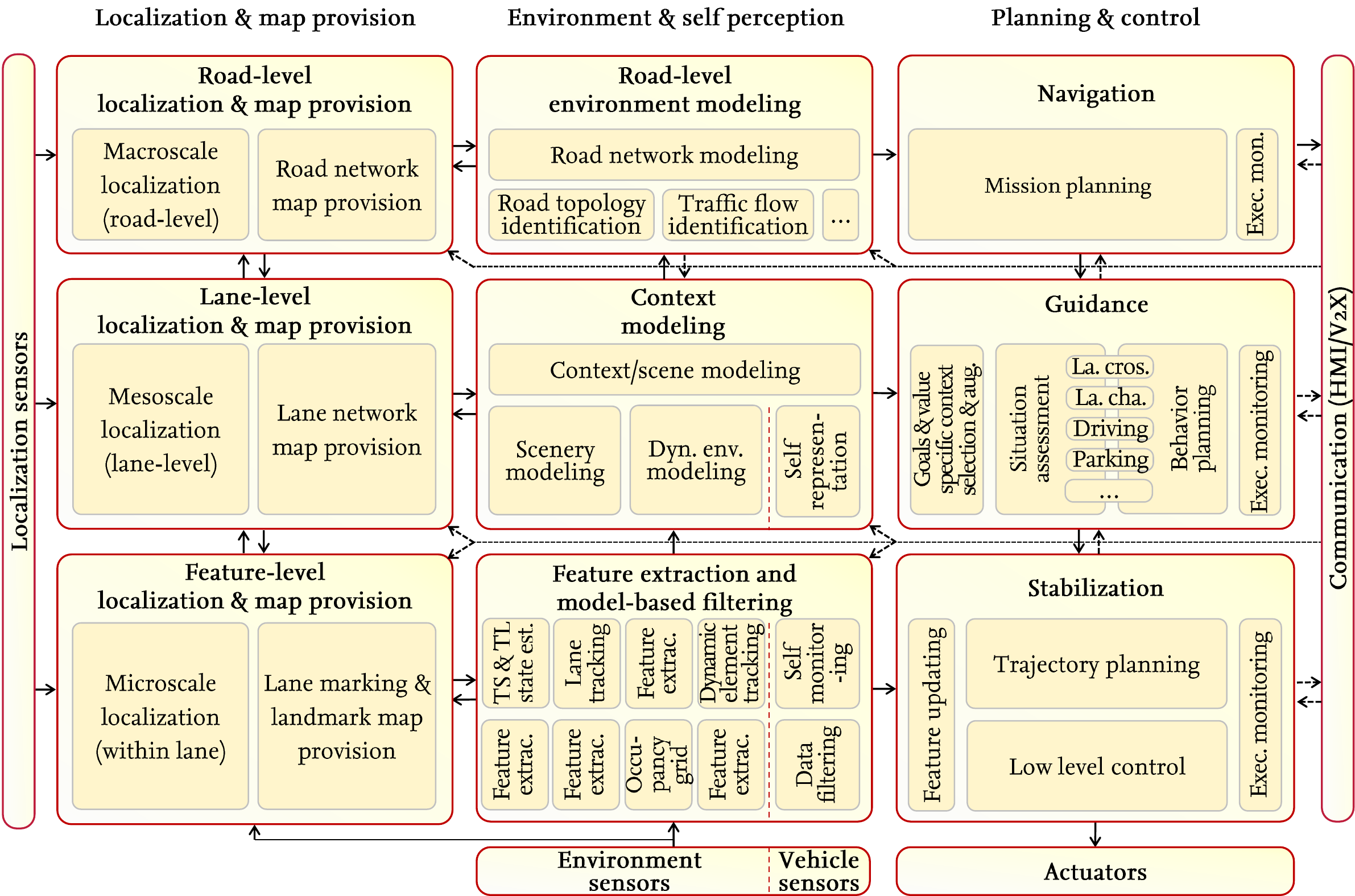}
			\centering
			\caption[Functional system architecture of an automated vehicle]{Functional system architecture of an automated vehicle. Blocks represent modules for activities; arrows show information flows (TS \& TL state est. = traffic sign \& traffic light (recognition and) state estimation, feature extrac. = feature extraction, dyn. env. = dynamic environment, aug. = augmentation, la. cros. = lane crossing handling, la. cha. = lane change handling,  exec. mon.= execution monitoring, HMI = human machine interface, V2X = Vehicle-To-X)}
      \label{fig:StadtpilotArchitecture}
			\vspace{-0.2cm}
\end{figure*} 

Transferring the concept of driving tasks and processing levels from human drivers to a technical system provides a starting point for a technical architecture.
In fact, this provides a hierarchical abstraction of driving tasks as in \citet{Maurer2000} and \citet{Matthaei2016}. Another distinction of tasks for automated driving may be derived from different processing steps of perceiving and acting.

Extending work by \citet{Zapp1988}, who described a functional control-cycle for automated vehicles, Hock \emph{et al.} \cite{Hock1992} showed an inverted \enquote{U} shaped signal flow from sensors to actuators with a hierarchical separation of processing levels for driving tasks.
Further specification and an exhaustive system description can be found in \citet{Dickmanns1994}. 
For instance in \cite[p. 239]{Dickmanns1988} and more clearly in \cite[p. 441]{Dickmanns2007}, Dickmanns presents the concept of separating \enquote{recognition} from \enquote{behavior (execution)} as well as the idea of aggregating features into abstract symbolic representations.
\citet{Maurer2000} and \citet[p. 185]{Dickmanns2007} highlight multiple feedback loops at different hierarchical levels constituting the signal flow in our  architecture nowadays.
In \cites{Dickmanns1988}[p. 595]{Dickmanns1996}[p. 387]{Dickmanns2007}, Dickmanns also illustrates the usage of a dynamic knowledge base and background knowledge that is now named as context\footnote{Context is here understood as the part of discourse that surrounds and represents an element.} modeling in this architecture.
Additionally, \cites[p. 40 ff.]{Maurer2000}[p.73 ff.]{Siedersberger2003}[p. 64]{Pellkofer2003}[p. 442]{Dickmanns2007} identify the central role of system capabilities and translate them into a hierarchical structure of abstraction.  
\cite[p. 442]{Dickmanns2007} separates between \enquote{scene understanding}, \enquote{planning}, and \enquote{gaze and locomotion control}.
As in \cite{Maurer2000}, the situation assessment is rather part of the perception.
In \cite{Dickmanns1994}, situation assessment stretches into both worlds: the planning column as well as the perception column.
In this article, the goal- and value-independent scene/context modeling is considered to be part of the perception column.
The goal- and value-specific situation extraction and situation assessment is considered to be part of the planning and control column.
The idea of \enquote{situation aspects} as a result of a \enquote{situation assessment} is presented in \cites[p. 51 ff.]{Pellkofer2003} for automated driving.   

\citet[p. 25\,ff.]{Matthaei2015a1}\marginnote{Literature Review} provides a more comprehensive literature review on the various forms of functional system architectures that have been used by different teams in automated driving as well as in robotics. For a broad literature review the reader is referred to his dissertation. Yet, we would like to relate our article to some recent publications on functional system architectures for automated vehicles.

\citet{Tas2016} compare the functional system architectures of several automated vehicles. They summarize the advantages and disadvantages of a distributed, modular architecture such as ours. They highlight the importance of fault detection, diagnosis, and self monitoring for system robustness in automated driving. Further, they unify the visual representation of three other vehicle architectures (\cite{Wei2014,Jo2015,Kunz2015}) into a common visualization scheme. Here, \cite{Tas2016,Wei2014,Jo2015} use a hierarchical structuring of driving tasks as in \citet{Donges1999} for \enquote{mission planning}, \enquote{behavior and motion planning}, and \enquote{vehicle control and actuation}. Perception, localization, and \enquote{vehicle state estimation} (cf. our self monitoring) are not hierarchically structured in \cite{Tas2016}, \cite{Jo2015}, or \cite{Kunz2015}. A \enquote{scene understanding} or \enquote{environment model} seems to be understood similarly to our context/scene modeling as a central point for information aggregation. \cite{Tas2016} suggests to consider Vehicle-To-X (V2X) communication as an \enquote{array of redundant sensors}. If such an approach is chosen, it is of particular importance to keep in mind that V2X information can be uncertain, incorrect, and even intentionally misleading. Further V2X communication might provide information on different levels of abstraction. Hence, we treat information from V2X differently than information from onboard sensors (cf. section \ref{sec:Communication}).

\citet{Behere2015} identify core components in a functional system architecture and group them under \enquote{perception}, \enquote{decision and control}, and \enquote{vehicle platform manipulation}. Their components resemble mostly our components. Yet, not all of our components are part of their architecture. Additional components they identified are \enquote{energy management} for \enquote{battery management} and \enquote{regenerative braking} and \enquote{reactive control} for reflex responses to unexpected stimuli as in automated emergency braking. For us, energy management is considered as part of the vehicle. Reactive control is indeed considered. It is part of the stabilization module and its low-latency data link explained in section \ref{sec:PlanningControl}. Like \cite{Tas2016}, \citet{Behere2015} consider V2X communication similar to a sensor/actor. Finally, \citet{Behere2015} identified the otherwise often neglected aspect of \enquote{diagnosis and fault management}. We agree to its importance. In our architecture it is part of \textit{self perception} (cf. section \ref{sec:ArchitecturePerception}) and \textit{execution monitoring} (cf. section \ref{sec:PlanningControl}). 

The functional system architecture presented here has been inspired by and applied to the Stadtpilot project for automated driving in urban environments and the aFAS project for an unmanned protective vehicle for highway hard shoulder road works. Hence, it is not just a top-down concept from a sketch board but has actually been proven to work in real world automated driving. It underwent several iterations. The foundations have been laid in \citep{Wille2010,Saust2010,Reschka2011,Wille2012}, the concept for context modeling has been developed in \citep{Nothdurft2011a,Ulbrich2014b,Matthaei2015b,Ulbrich2015d}, environment perception has been refined in \citep{Rieken2015,Matthaei2015b,Matthaei2015a1}, self representation has been addressed by \citep{Reschka2015,Stolte2015,Reschka2017}, and localization and map provision has been discussed by \citep{Matthaei2014,Matthaei2015b,Matthaei2015a1}. The remainder of this article will show the status quo of our functional system architecture and present the enhancements compared to previous publications.

\section{Functional System Architecture}
Figure\marginnote{Foundations} \ref{fig:StadtpilotArchitecture} illustrates a revised architecture 
based on \cite{Dickmanns1994,Dickmanns2007,Maurer2000,Wille2010,Saust2010, Reschka2011,Nothdurft2011a, Wille2012,Ulbrich2014b, Rieken2015,Matthaei2015b,Matthaei2015a1}.

The\marginnote{Vertical Abstraction Layers} vertical abstraction layers of the functional system architecture are aligned to the levels of driving tasks from \citet{Donges1999}, \citet{Riemersma1979}, and \citet[p. 498]{Michon1985}. One elementary (operational) stabilization layer is used for course keeping and speed control, a second (tactical) guidance layer is for any behavior planning and a third one is for strategic planning (navigation). \citet[p. 281]{Albus1979} suggested the use of such a hierarchical structure not only for behavior planning and control but also for perception. \citet{Nothdurft2014} transferred the concept of \citet{Oberlander2008}, to differentiate context information in particular for digital maps\footnote{Based on \cite{OxfordMap} and \cite{Ulbrich2015d}, we understand a map as a diagrammatic representation of an area's scenery.} by \enquote{topological,} \enquote{semantic,} and \enquote{metric} properties to the field of automated driving. In Figure \ref{fig:StadtpilotArchitecture}, the terms road-level relate to the road network topology, lane-level to the semantic relationships among lanes and feature-level to the metric properties used for a localization within a lane.

Certain modifications have been made by the team at the Institute of Control Engineering at TU Braunschweig after the publication of previous architecture versions in \citet{Matthaei2015b} and \citet[p. 37\,ff.]{Matthaei2015a1}. The following sections describe the current state of the functional system architecture and discuss the recent modifications. It is explained along the inverted-\enquote{U}-shaped main signal flow through the components in the architecture. 

\subsection{Environment and Self Perception}
\label{sec:ArchitecturePerception}
\subsubsection{Interfaces}
The environment and self perception column has interfaces to localization and map provision, behavior planning and control, and with the communication column. Within the column there is an interface towards sensors.

Perception\marginnote{Sensors} has an interface towards the automated vehicle's sensor systems. They have been clustered into environment sensors covering external aspects around the vehicle (exteroceptive) and vehicle sensors to obtain information about the vehicle itself and its internal state (proprioceptive). Environment sensors are sensors like cameras, lidar, and radar sensors but also conventional sensors like a thermometer or a rain sensor. Vehicle sensors provide information about the movement or pitch of the ego vehicle, but also information about the charging/filling level of the battery/the fuel tank, for example.
In a hardware architecture, sensor data feature extraction and even model-based filtering may be allocated to a sensor itself. Yet, in a functional system architecture, the interface between the sensor block and the subsequent feature extraction is raw sensor data.  

Although\marginnote{Inputs} the perception column is primarily based on sensor data from within the column, it may use map information together with a pose within that map as input on different hierarchical levels of abstraction. On a macroscale level, there are topological road network maps used to augment perceived information with a-priori map information. On a mesoscale level, lane level map information may be used to augment context modeling even beyond the limited field of view from on-board sensor systems. On a microscale level within a lane, feature information may be used to provide additional landmarks or to stabilize lane tracking. Likewise an input might be Vehicle-To-X information obtained from other traffic participants or infrastructure.  

The algorithms in the planning and control column are the primary\marginnote{Outputs} data user of the perception column. On a navigation level, a road network together with a traffic flow may be used to calculate an optimal route. At the tactical level (guidance), a scene\footnote{A scene describes a snapshot of the environment including the scenery and dynamic elements, as well as all actors' and observers' self-representations, and the relationships among those entities. Only a scene representation in a simulated world can be
all-encompassing (objective scene, ground truth). In the real world it is incomplete, incorrect, uncertain, and from one or several observers' points of view (subjective scene) (cf. \citet{Ulbrich2015d}).} as defined in \citet{Ulbrich2015d} is provided. On an operational level (stabilization), the perception may provide simple features and state variables as a low latency shortcut to low level control as in \citet[p.~42]{Maurer2000}.

Perceived information is provided on different levels of abstraction (road-level, lane-level, feature-level) for map updates or mapping. Sensor data (gyroscopes, wheel tick sensors,~...) from the perception column may directly be used for localization and map provision. 

Last of all,\marginnote{Communication} perception data may directly or indirectly be used for broadcasting information via Vehicle-To-X communication or visualization. The authors assume that there will always be a goal and value specific context selection. Thus, rather a for others as \textit{relevant} classified situation subset will be communicated or visualized. Yet, also with this intermediate step, communication will be at least \textit{based} on information from perception.
 
\subsubsection{Comprised Activities}
\begin{figure*}[htbp]
      \centering
			\includegraphics[width=\textwidth]{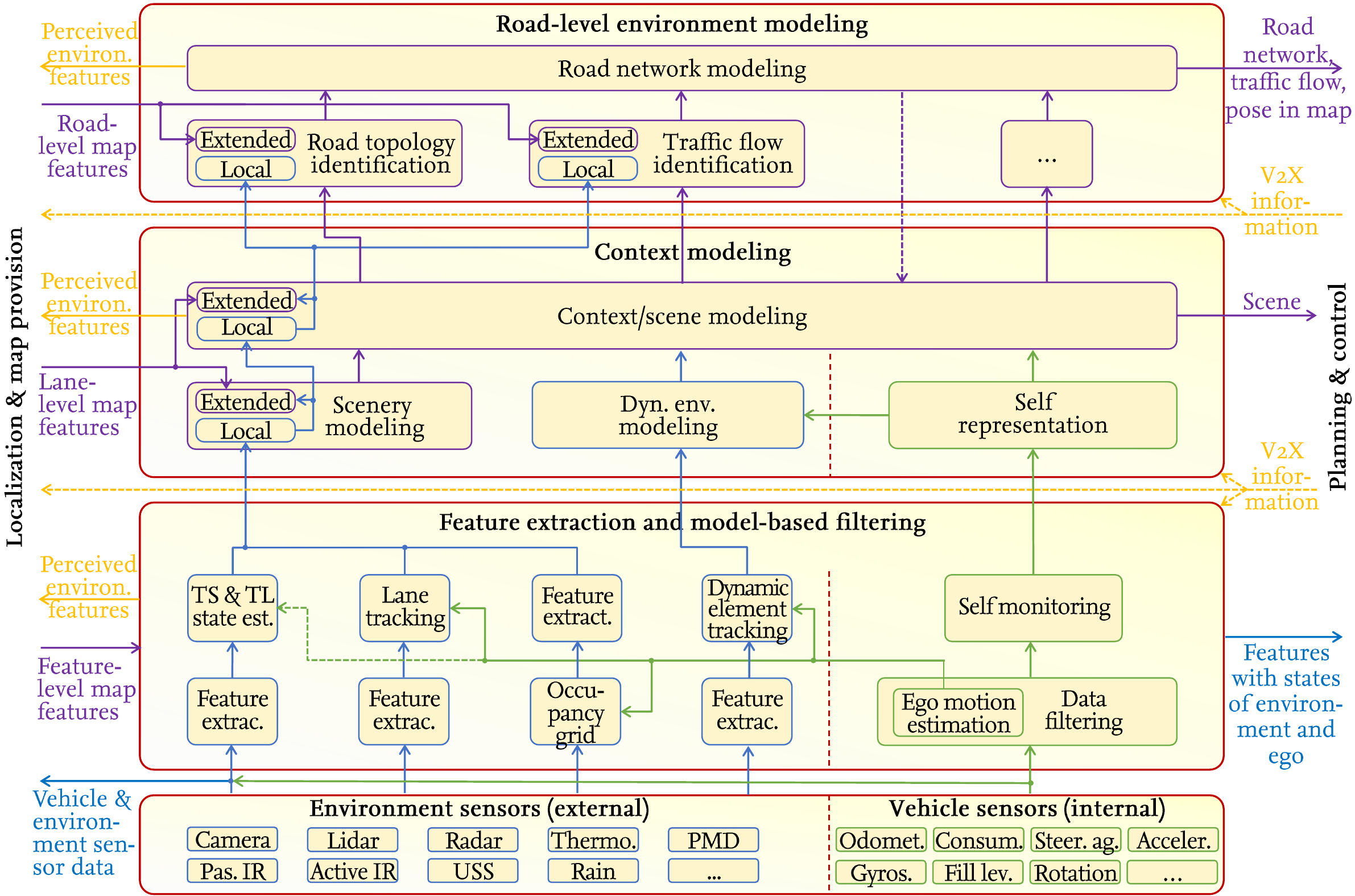}
			\centering
			\caption[Environment and self perception]{Environment and self perception based on \citet{Rieken2015} and \protect\citet{Matthaei2015b}. Green = only subject to vehicle sensor errors; blue = subject to environment sensor errors; violet = also subject to map and localization related errors; yellow = perceived environment features and Vehicle-To-X information (dyn. env. = dynamic environment, TS \& TL state est. = traffic sign \& traffic light (recognition and) state estimation, extrac. = extraction, pas. = passive, IR = infrared sensors, USS = ultrasonic sensors, thermo = thermometer, PMD = photonic mixing device, rain = rain sensors, odomet. = odometry sensors, gyros. = gyroscope, consum. = consumption sensors, fill lev. = fill level sensors, steer. ag. = steering angle sensors, rotation = rotation sensors, acceler. = acceleration sensors, environ. = environment, V2X = Vehicle-To-X)} 
      \label{fig:Perception}
			\vspace{-2mm}
\end{figure*}

Figure \ref{fig:Perception} provides details on the environment and self perception. The dashed line symbolizes the separation between the perspective to the outside (environment perception) and the often neglected perspective to the inside (self perception) as in \citet[p. 58\,ff.]{Maurer2000}, \citet[p. 145\,ff.]{Bergmiller2015}, and \citet{Reschka2015}. Similar as in \citet[p. 51]{Matthaei2015a1}, a green color codes that only relatively certain internal information has been used. The blue color indicates that only internal sensors \textit{and/or} environment sensor information has been used. The violet color indicates that additionally map data with all possible errors in map-relative localization and incorrect, possibly outdated map information has been used. The yellow color indicates perceived data used for map updates and Vehicle-To-X information.

Sensor data is used for feature extraction and subsequent model-based filtering. Feature\marginnote{Feature Extraction and Model-Based Filtering} extraction and model-based filtering is performed regarding several aspects. This includes lane detection and tracking, dynamic element tracking\footnote{\enquote{Dynamic objects} form the set of \enquote{dynamic elements} by extending them with non-object-model-compliant elements (cf. \citet{Ulbrich2015d}).}, occupancy grid modeling plus subsequent feature extraction and data filtering, traffic sign and traffic light recognition and state estimation, as well as self monitoring of the automated vehicle. Input to this block are raw or processed sensor data and possibly feature-level map data. Models are used to identify entities, associate measurements to entity hypotheses and track entities over time. In lane tracking, dynamic element tracking, and traffic light and traffic sign recognition a temporal validation or tracking is typically performed after an extraction of relevant features. In occupancy grid mapping, widely used for the stationary environment, a similar temporal filtering results from a probabilistic filtering performed in different cells of the occupancy grid itself. Entities and properties of these are generated by a subsequent feature extraction from that grid.\footnote{The feature extraction and model-based filtering is not discussed in further detail here. Some details are provided in \citet{Rieken2015} and will be discussed in a future publication specifically on this topic.}

Any of\marginnote{Self Perception} the sensors are mounted to the automated vehicle. Thus, their sensor data will be ego-relative. To transform sensor data into a stationary coordinate system, it is necessary to estimate ego motion. This is part of the data filtering in self perception. We further suggest to integrate self monitoring into self perception. The threshold between a self monitoring and self representation on a context modeling level seems vague at first. The authors suggest to use the same differentiation as for other entities. The self monitoring provides information about entities of the ego vehicle and their attributes like health states or errors. The self representation provides semantic links between those entities to derive a full context not only about the environment but also about the automated vehicle itself. 

The\marginnote{Context/ Scene Modeling} information from the feature extraction and model-based filtering is used for context/scene modeling (cf. \citet{Ulbrich2015d}). This subsumes several aspects of information modeling, aggregation, and association. Scenery modeling combines lane information with a scenery model. This scenery model may use a-priori map data and a position in this map from the localization and map provision column in Figure \ref{fig:StadtpilotArchitecture}. Dynamic environment modeling may interact with the scenery model to incorporate model-based information. Dynamic elements, for example, are more likely to move along lanes or paths.\footnote{For safety applications and to model non-rule compliant behavior, it is essential that this is only an information augmentation. The initial tracking results still need to be maintained to avoid crashing into non-rule compliant dynamic elements.} Dynamic elements and the scenery are associated with each other to obtain an environment model. This is combined with the self representation of the ego vehicle to yield a context/scene model. This scene representation is transmitted to modules in the planning and control column. \citet[p. 52]{Matthaei2015a1} differentiates a \enquote{local} scenery and scene modeling from an \enquote{extended} one. The first is solely based on perceived information and incorporates no map-related information. Its output can be used for updating a map with perceived information. The distinction avoids loops in the information flow and self-confirming hypotheses of confirming map data with map-supported perception data. 

The perception\marginnote{Road Network} column is completed by modeling a road-level environment. This subsumes a possible road topology and traffic flow identification to estimate which lanes constitute roads and whether these roads are congested or blocked.\footnote{A lane level traffic flow identification may still be considered as part of the context modeling.} So far, this module has not been implemented in the Stadtpilot or aFAS project. The road network is simply piped through as it is from an a-priori map from the localization and map provision column towards subsequent modules.

\subsubsection{Enhancements to the State of the Art}
The modifications\marginnote{Feature Extraction \& Model-Based Filtering} are shown towards \citet{Matthaei2015b} as the last broadly accessible publication of our functional system architecture in English. The sensors' block is identical; feature extraction and model-based filtering has only been marginally modified regarding the self perception. Here, Matthaei only mentioned the aspect of \enquote{motion estimation} and a rather vague \enquote{data filtering} (\citeauthor{Matthaei2015b}, \citeyear{Matthaei2015b}, p.~162; \citeauthor{Matthaei2015a1}, \citeyear{Matthaei2015a1}, p.~51). Yet, as in \citet[p. 58\,ff.]{Maurer2000}, \citet[p. 145\,ff.]{Bergmiller2015}, and \citet{Reschka2015} this is only part of the self perception. It may further include friction coefficient estimation, vehicle component wear-and-tear estimation, component diagnosis, energy level estimation, etc. 

\begin{figure*}[htbp]
      \centering
			\includegraphics[width=\textwidth]{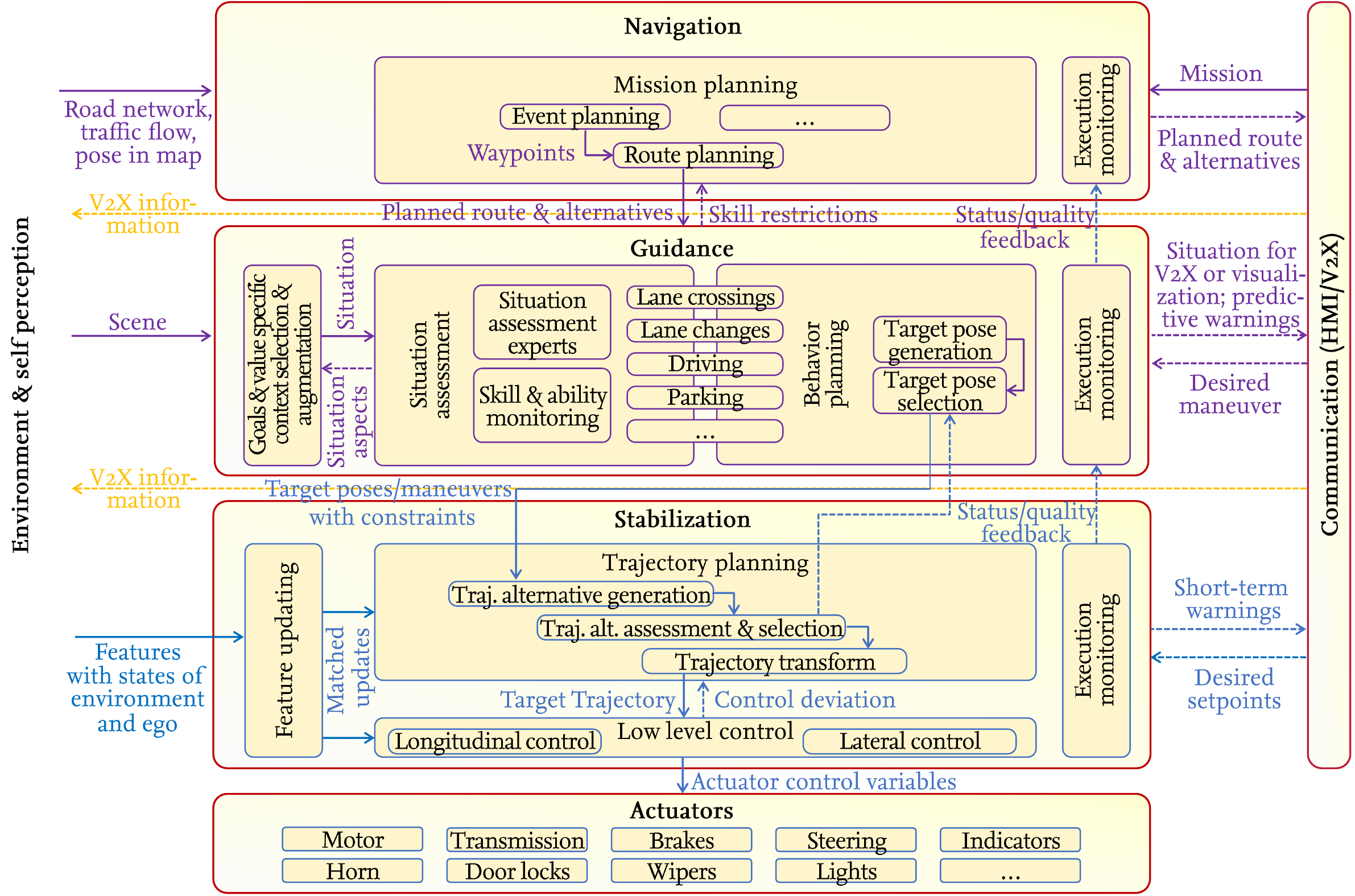}
			\centering
			\caption[Planning and control]{Submodules for planning and control. Blue = subject to environment sensor errors; violet = also subject to map and localization related errors; yellow = Vehicle-To-X information for perception (V2X = Vehicle-To-X, dyn. = dynamic, traj. = trajectory, alt. = alternative, HMI = human-machine-interface)}
      \label{fig:Planning}
			\vspace{-0.2cm}
\end{figure*}

The aspect of traffic sign and traffic lights has marginally been modified. \citet[p. 162]{Matthaei2015b} called it traffic sign and traffic light \enquote{detection}, \citet[p. 51]{Matthaei2015a1} called it traffic sign and traffic light \enquote{state estimation}. Of course it is necessary to detect, recognize the position, type and in case of traffic lights the state of an element. Other than tracking stationary lane markings/lanes, Matthaei assumes no need for an ego motion compensation for traffic signs and lights. Traffic sign and traffic light estimation has not been implemented in the Stadtpilot project. If this is purely frame-based, it may indeed not need an ego motion estimation. If it stabilizes traffic sign/light hypothesis over time, it will need an ego motion. Thus it has been linked by a dotted line.

Context\marginnote{Context Modeling} modeling has been restructured. 
Matthaei's differentiation between \enquote{local} scenery/scene modeling and \enquote{extended} scenery/scene modeling\footnote{This entails information from a-priori map data about static and movable elements.} have been both subsumed under only one scenery modeling and scene modeling with corresponding submodules. A dynamic environment modeling has been introduced as an analogon to scenery modeling for static environment aspects. This may include steps of validating different tracks of dynamic elements against each other. For instance if the contours of different elements overlap it might be a sign of actually tracking the same object twice rather than in fact observing a collision. 
Further, \citet[p. 162]{Matthaei2015b} and \citet[p. 51]{Matthaei2015a1} called the step of associating semantic information about the automated vehicle \enquote{vehicle state modeling}. Aligned with \citet[p. 145\,ff.]{Bergmiller2015}, the authors prefer \textit{self representation} as a name for this block. 
Last of all, the name of the overall module seems odd at first. While it is named \textit{context modeling}, its output is only a \textit{scene} from a \textit{scene modeling} as in \citet{Matthaei2015b}. With the definitions from \citet{Ulbrich2015d} it is indeed correct to have a scene as an output. Yet, the process itself entails aspects of context modeling, too. Thus the name of the module is extended to \textit{context/scene modeling}. 

Similar\marginnote{Road-Level Environment Modeling} to \citet[p. 51]{Matthaei2015a1}, road topology identification and modeling as well as traffic flow identification are summarized in a block above context modeling. The block has been renamed from \enquote{road topology and traffic flow modeling} to a more general road-level environment modeling. Linguistically, this makes room to identifying and modeling one day aspects like ferryboats affecting the mission planning due to limited operating hours as a part of this block. Moreover, an arrow between road-level environment modeling and the context/scene model has been added to represent such an information flow of high-level road information towards information in a scene.

\subsection{Planning and Control}
\label{sec:PlanningControl}
\subsubsection{Interfaces}
The planning and control column has interfaces to the perception and communication column and towards the actuators within the column.
Inputs from perception are:\marginnote{Inputs} 
\begin{itemize}
	\item A road network together with a traffic flow information for navigation.
	\item The scene described as in \citet{Ulbrich2015d} for tactical planning (guidance).
	\item Features with state variables as a low latency shortcut to control as in \citet[p. 42]{Maurer2000}.
\end{itemize}

Outputs\marginnote{Outputs} exist within the column towards actuators. These entail gas, gears, brake commands, and steering. Yet, it may also include actuation of other vehicle components like the horn, indicators, or headlights. It may even include opening a door lock or the trunk for freight delivery or loading, or activating the wipers for removing dirt from the windscreen.

Interfaces towards the communication column will be detailed in its according section.

\subsubsection{Comprised Activities}
Figure \ref{fig:Planning} illustrates details on the planning and control in a functional system architecture. The color coding for information flows is the same as in section \ref{sec:ArchitecturePerception}. Modules\ for planning and control use the previously mentioned scene as a central interface on a tactical level. The modules have been divided into three levels according to the hierarchy of driving tasks in \citet{Donges1999}. 

On\marginnote{Navigation} a strategic level (navigation), the road network, information about traffic flows or blockages and an externally provided mission are used for navigation purposes. A mission planning as in \citet[p. 405]{Dickmanns2007} or \citet[p. 81\,ff.]{Gregor2002} entails planning certain events like a cargo or passenger pickup. They result in waypoints between which a route needs to be planned. A route planning yields a ---with respect to some optimization criteria--- best route but also route alternatives.\footnote{The aspect of route alternatives has so far not been implemented in the Stadtpilot or aFAS project.} The calculation of route alternatives may be triggered by events or upon request of the tactical level (guidance). The navigation may consider skill restrictions of an underlying guidance layer. If, for instance, the battery of an electric vehicle is too low to take a shorter but more energy consuming route through a mountain area. Route alternatives to reach the mission goals are recalculated to reflect ego position changes.
 
The guidance modules use this information to render a mission executable.\marginnote{Guidance} They use the current scene to select relevant aspects and to augment it with additional information to derive one or several situation representations for the automated vehicle. Such a situation is used for situation assessment and behavior planning regarding several situation aspects. Among those are regular driving within a lane, lane changes, lane crossings (e.g., at intersections), free space navigation for parking, etc. (cf. \enquote{driving maneuvers [...] for automated vehicles}, \cite[p.~122\,ff.]{Reschka2017}). 
Situation assessment for these situation aspects entails application specific situation assessment expert algorithms and also skill and ability monitoring for that particular situation aspects. Behavior planning entails not only maneuver selection but also planning about how a maneuver should be executed. This \textit{how} does not include detailed velocity profile planning but rather a sequence of tactical behavior decisions like longitudinal and/or lateral adjustments to a gap or stopping points in an intersection, indicator activations or maybe even honking one's horn. The guidance block is completed with execution monitoring of all components, which ensures reliability (continuity of correct service) and availability (readiness for correct service) \cite{ISO24765}. This execution monitoring has ultimate control over deactivating the system or its modules.

Output\marginnote{Interface between Guidance and Stabilization} of the tactical guidance layer is a set of target poses for maneuvers. A target pose commands the stabilization layer \textit{what} to plan for. This may entail a target position, orientation, velocity (and further derivatives), constraints for trajectory planning like a drivable area, a reference corridor, sampling ranges or target deviation costs, and a symbolic maneuver type information.  

The maneuver information may be utilized by an underlying stabilization level to switch between algorithms as in \citet[p. 74]{Maurer2000}. 
The target pose may be linked to a vehicle with a certain id to perform longitudinal vehicle following. 
It may be set to the center of a neighboring lane for lane changing or it is set towards a gap in traffic for longitudinal adjustments to prepare lane changing. 
For parking, this pose may contain a goal position and orientation in a parking lot. 
Even at complex intersections, this interface seems sufficient to cover, e.g., to stop at a stop sign, proceed to a line of sight and finally turning through a lane with oncoming traffic. 

Depending on the actual implementation only one or several\footnote{So far, only one target pose has been implemented in the Stadtpilot or aFAS project.} target poses may be handed over to the stabilization level. In case of the latter, target pose selection is implicitly done by the knowledge of selection rules in the stabilization level. 

The stabilization subsumes trajectory planning and low level control, and execution monitoring as three major aspects.
Trajectory planning calculates trajectory candidates for all these target poses. Low level control translates those trajectories into actuator control variables. Execution monitoring detects deviations between what is planned and executed.

Trajectory planning\marginnote{Trajectory Planning} as in \citet{WerlingDiss2010}, or path planning with a subsequent velocity profile planning as in \citet{Kammel2008}, \citet{Hundelshausen2008}, \citet{Wille2012}, \citet{Broggi2013} can be generalized into a three step procedure of trajectory alternative generation, trajectory alternative assessment and selection, and transforming the results into a representation to be used for low level control.

A path or trajectory planning may entail a subsampling of further target poses around the provided target poses as in \citet[p. 42]{WerlingDiss2010}. Based on a cost function, the best trajectory, according to a cost criteria, is selected.\footnote{Selecting the best point could once more be considered as tactical decision making. Hence, one could argue the necessity of a trajectory selection arbiter block within the guidance module. For simplicity, it is excluded in Figure \ref{fig:Planning}.} 
Depending on the implementation of the trajectory planning, it is necessary to transform the trajectory from a geo-stationary, local coordinate system of a scene or situation towards an ego-vehicle bound coordinate system in which the actuators and low level controllers operate. If trajectory planning is executed in a Frenet frame, this transform is performed as a last step.

A\marginnote{Low Level Control} future point on this trajectory is used as input to the low level controllers to command a steering angle, brake pressure, or acceleration rate to the actuators of the automated vehicle. To reduce latency, it may be necessary to obtain direct feature updates from the previously mentioned model-based filtering algorithms directly on the stabilization level. These feature updates may be incorporated into the low level controllers or even the trajectory planning. 

Once\marginnote{Execution Monitoring and Feedback} more, the stabilization level entails execution monitoring to ensure the correct functioning of these algorithms and possibly to inform the guidance module about issues on the stabilization level. Examples of this driving task relevant information are if no collision-free trajectory can be calculated or if the execution of commanded behavior is not possible due to physical limitations in the vehicle's dynamics. This feedback is either used for execution monitoring in the tactical level or even to adopt the tactical behavior planning (guidance) or strategic mission planning (navigation). For instance, if changing lanes to a highway exit lane jam-packed with traffic requires high relative velocity adjustments and thus high discomfort in trajectory planning, it may even affect the route planning by avoiding such a maneuver and simply taking an alternate route by choosing a next exit further down the highway. Likewise, even low level control may provide such feedback by reporting control deviations. If a high slip angle indicates issues in vehicle stability, it may even affect tactical behavior planning by changing to a lane with better friction.

\subsubsection{Enhancements to the State of the Art}
On the strategic level of navigation, the route planning has been renamed to a more general mission planning. When the scope of automated driving becomes wider, mission planning may not only contain route planning but even mission elements \cite[p. 43]{Gregor2002a} like cargo pickup, or refueling.
\citet{Matthaei2015b} mention a \enquote{selection of a next navigation point} as a submodule of the navigation block. Only transferring the next navigation point to a guidance module imposes a severe limitation because several route alternatives may exist. 
This can be illustrated in the earlier mentioned example of an automated vehicle performing a lane change onto an off-ramp jam-packed with traffic. If there is a high risk to exceed the skills of the vehicle, it may be better to avoid such a risky lane change and accept a marginal detour rather than to enforce exiting where it was planned.
 
This is not only a thought experiment but rather a real world issue and addressed by the lane advice in \citet{Ulbrich2015b}. For that reason, the authors deviate from \citet{Matthaei2015b} by assuming not only one but several routes as an output of the route/mission planning and dropping the \enquote{selection of a next navigation point} altogether. Only if the alternatives are known, an informed tactical decision about following or deviating from what was planned at the navigation level is possible.
Likewise to incorporate such knowledge about limited skills from a tactical level (either from the self representation as part of the scene) or the situation assessment and behavior planning itself into the mission planning, an upward facing arrow from guidance to navigation is added.

Deep changes have been made to tactical planning compared to \citet{Matthaei2015b}. As illustrated in \citet{Ulbrich2015d} a goals- and value specific context selection and augmentation is added as an intermediate step between a goals- and value independent scene and a goals- and value related situation. There may be one or several situation data structures for different aspects of behavior planning. 
They can be used as an input or even be augmented by modules for situation assessment.\footnote{Other than \citet{Matthaei2015b} the authors prefer the less ambiguous term situation assessment instead of situation analysis. Yet, a situation is rather the input of a situation assessment than its output. Only some situation aspects may be needed for other modules in situation assessment and thus fed back into the situation data structure.} For instance, the results of a gap quality assessment might be fed back into a situation. That information could be used in an adaptive cruise control target pose selection module to temporarily reduce a time gap towards a front vehicle to avoid restricting gap adjustments to a gap slightly in front.  

Behavior planning is used as an additional block to reflect not only a maneuver selection but likewise the earlier introduced planning about \textit{how} a maneuver should be executed. The earlier introduced execution monitoring is added as an additional block to the planning and control column. No clear opinion has yet been formed if it is\ignore{actually} necessary to include execution monitoring as a separate block or if every block is supposed to have a sub-aspect of execution monitoring.
Yet, as mentioned earlier, it is indeed important to include the \textit{upward} information flow from stabilization to guidance.\footnote{This extension is based on discussions with Professor Chris Gerdes, Stanford University in 2014.} It was missing in \cite{Matthaei2015b} and has now been added.

The\marginnote{Stabilization} stabilization block has been detailed compared to \citet{Matthaei2015b}. A feature updating block has been added to reflect the updating process of, e.g., vehicle distances and velocities for low latency stabilization (cf. \cite[p. 42]{Maurer2000}). Trajectory target poses from the guidance level may be associated to dynamic elements. Their dynamic state variables may be updated based on more recent information directly from model-based filtering while bypassing the latency induced by the more comprehensive context modeling, situation assessment, and behavior planning. This leads to faster reactions in time critical scenarios. 

The\marginnote{Actuators} set of actuators has been extended by adding indicators, the horn, door locks, wipers, lights, etc. \citet[p. 164]{Matthaei2015b}\ignore{\citet[p. 58]{Matthaei2015a1}} highlight that some actuators are used for the purpose of tactical communication (cf. \enquote{implicit communication} in \citet{Ulbrich2015c}). These actuators (or rather: devices) have not been part of the functional system architecture so far, neither as part of the communication column nor of the actuator block. Due to their similar nature as activating a brake light, they are all grouped under the actuator module. A module from the tactical guidance level may actuate those devices \textit{through} the operational stabilization level.

At last, the \enquote{planning and control} column has been renamed from the linguistically ambiguous\ignore{vague} term \enquote{mission accomplishment}.
 
\subsection{Communication}
\label{sec:Communication}
\subsubsection{Interfaces}
The\marginnote{Strategic Level} interfaces of the communication are illustrated in Figure \ref{fig:Planning}. At the strategic level for navigation tasks, a mission may directly be commanded from an operator via a human-machine interface or even remotely via Vehicle-To-X communication. The mission may entail a route destination as well as goal criteria like a route with most comfort in automated driving, shortest travel distance, or the most economic route alternative.
As a feedback, the system may communicate a planned route, resulting from the commanded mission. Yet, the system may even provide route alternatives to an operator to enhance mission selection. The authors agree with \citet[p. 56]{Matthaei2015a1} that for a SAE-level-5 system (cf. \cite{SAE2016}) of an automated vehicle, the only necessary input is on a strategic level (navigation). Yet, for the sake of informing an operator or in case of not-level-5 systems, additional communication interfaces are necessary.

At\marginnote{Tactical Level} a tactical level for guidance tasks, a situation is used as an interface for visualization and Vehicle-To-X communication. 
While the situation for Vehicle-To-X communication may be different from the situation for behavior planning of the ego vehicle, it is still a situation because not every aspect that is part of the scene will be \textit{relevant} for the (assumed) goals and values of any of the information recipients in Vehicle-To-X communication, or \textit{legal} to be transmitted (cf. \enquote{enhancements} section).
Likewise, a situation for visualization will probably be simplified and temporarily smoothed to reduce distraction. Yet, it is still a situation because it shows what is relevant regarding the goals and values of an operator or interested passenger. It may entail information about planned maneuvers as part of the situation aspects derived from planning and control. \textit{Predictive warnings} to inform a passenger may either be considered as part of the situation or as a separate information interface from the guidance module towards the communication column.

In the opposite direction (towards perception and map provisioning), the communication column provides Vehicle-To-X information to be incorporated into the scene and possibly likewise on a feature or road level. Likewise, a desired maneuver may be commanded from an operator to the guidance module \cite[p. 57]{Matthaei2015a1}. This could be to command an operator-initiated lane change but also to command an emergency stopping maneuver or a driver takeover request.

At\marginnote{Operational Level} the operational level (stabilization), short term warnings may be issued or desired setpoints commanded \citep[p. 57]{Matthaei2015a1}. Short term warnings could be the activation of an electronic stability control system in case of a higher than intended slipping angle on a low friction road. A desired setpoint could be the timegap towards a leading vehicle for an adaptive cruise control driver assistance system. For a future \mbox{level-5} system these interfaces may not be necessary anymore, because by definition the system needs to handle all these aspects without driver intervention. Yet, as long as there is a transition between humans used to drive a vehicle by themselves and full automation these interfaces may still exist as a legacy for a long time. 

\subsubsection{Comprised Activities}
An automated vehicle may have a communication interface for communicating with an operator or passenger (human-machine interface, HMI), as well as for technical communication with other traffic participants or the infrastructure via a Vehicle-To-X (V2X) communication interface. 

The human-machine interface\marginnote{Human-Machine Interface} entails both directions of communication: On the one hand, to obtain input from an operator or passenger and on the other hand to provide information. A special case are automated vehicles being monitored by a central tele-operation unit. Here the aspect of a human-machine interface and the usage of communication networks are combined. \citet[p. 56]{Matthaei2015a1} envisions the idea of navigation or guidance inputs for traffic management or clearing corridors for emergency vehicles. 
For the latter, the reliability and guaranteed coverage of current communication networks is an issue. Yet, at least the technically less demanding centrally controlled deactivation of an automated driving function within a certain amount of hours could be useful to ensure the absence of hazardous states caused by a bug, after such a bug has been discovered in the fleet of automated vehicles. 
  
The\marginnote{Vehicle-To-X Communication} aspect of Vehicle-To-X communication entails communication with other traffic participants or infrastructure. Depending on what other vehicles are able to provide the range of applications is wide. Current research initiatives like Ko-HAF\footnote{http://www.ko-haf.de/, visited on Nov. 29th, 2016.} address aspects like obtaining map updates from fleets, collaborative perception, and coordinating cooperative driving maneuvers among traffic participants. Algorithms to implement such behavior are spread among the modules in the other three columns of the functional system architecture. Yet, the actual communication interface for 802.11p wireless local area network communication, cellular network communication, or other communication channels is part of this column.  

\subsubsection{Enhancements to the State of the Art}
Certain modifications have been made to the communication column since it was published in \citet{Matthaei2015b}.  

Regarding interfaces, changes have been made to some contents of existing arrows. The interface between navigation and communication in \citet[p. 57]{Matthaei2015a1} is extended by not only exchanging a \enquote{route} but rather a \enquote{mission} as input to the navigation and by adding the aspect of route alternatives for the opposite information flow. 

While Matthaei\footnote{Internal report \enquote{Cooperation, Collaboration, and Communication} from March,  2015.} assumed collaboration happens over the interface left of the perception column, the authors suggest to use the existing communication interface in the communication column. To the authors, there is no need for a separate interface in the functional architecture, because aspects from the perception column can be exchanged with one interface at the very left. 
To allow an information flow from the communication column to the perception and localization and map provision columns, additional links have been added. 

For transmitting Vehicle-To-X information, it is assumed that a full scene will probably never be sent but only a \textit{relevant extract} of the aspects assumed to be relevant for the information recipients and their archival of their anticipated goals and values (situation for Vehicle-To-X communication). If little information exists about the goals and values of the information recipients, only obviously irrelevant aspects (e.g., privacy, what was seen inside of buildings by accidentally looking through windows) may be excluded and thus the \textit{relevant extract} may almost converge against the full information from a scene.\footnote{In the distinction between a scene and situation in \citet{Ulbrich2015d} the focus was rather on goals and values of a vehicle. Here the distinction has similarly been extended towards goals and values to be considered for communication as they are stipulated by authorities, e.g., privacy.}
If legislation and communication channel width will ever allow to broadcast a full scene, the aspect of information selection could be dropped and the link between the perception and communication blocks becomes bidirectional.

The localization and map provision column can exchange V2X information with the communication interface. 
Thus, the blocks in localization and map provision can receive and send updates of map data on all layers of the architecture. 

\subsection{Localization and Map Provision}
\subsubsection{Interfaces}
The localization and map provision column has interfaces with the perception column to exchange:
\begin{itemize}
	\item road-level map features and map updates,
	\item lane-level map features and map updates,
	\item feature-level map features and map updates, and
	\item vehicle and environment sensor data.
\end{itemize}

Further,\marginnote{Localization Sensors} it has an interface with localization sensors. According to \cite{Matthaei2015b}, the localization sensors like those in a global navigation satellite system (GNSS) are not part of the environment sensors at the bottom but are rather noted on the left due to providing information on higher abstraction levels. 

Within\marginnote{Within the Column} the column, information is exchanged between the different hierarchical levels. The upward information flow represents the use of, e.g., low level map features to extract higher level lane information. Likewise, there is an information flow downwards: Information about the existence of a road might be used to establish semantic relationships and support lane hypotheses in a lane level map. 

\begin{figure*}[htbp]
      \centering
			\includegraphics[width=\textwidth]{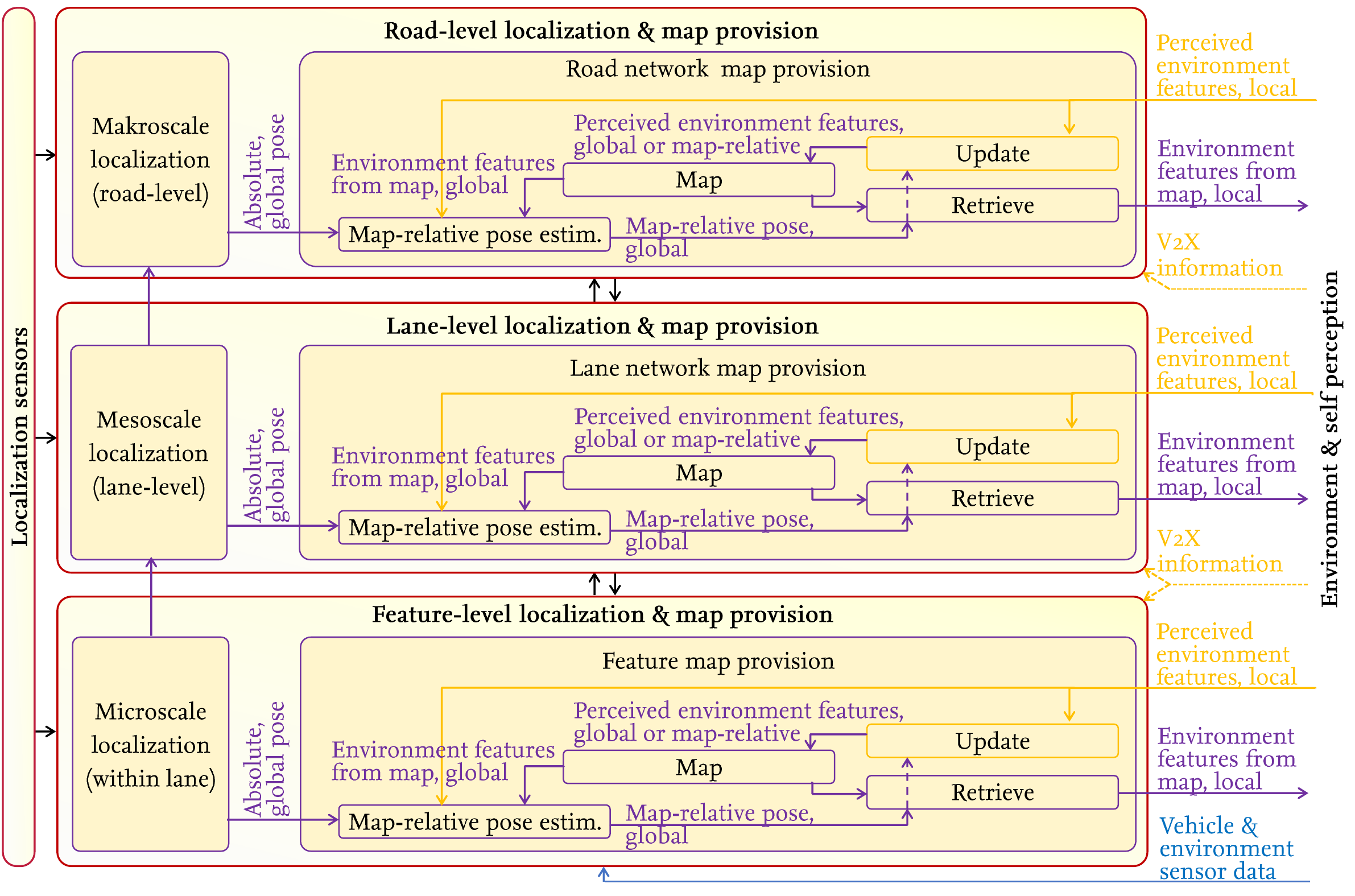}
			\centering
			\caption[Localization and map provision]{Localization and map provision based on \protect\citet[p. 45]{Matthaei2015a1}. Blue = subject to environment sensor errors; violet = also subject to map and localization related errors; yellow = features for map updates (estim. = estimation)} 
      \label{fig:Localization}
			\vspace{-2mm}
\end{figure*}

\subsubsection{Comprised Activities}
The automated vehicle needs to localize itself relative to its maps to make use of information in these maps. The aspect of map provision entails providing map information to other modules as well as the process of mapping and map updating in order to have such information to share. All these aspects are depicted in Figure \ref{fig:Localization}. 

Localization\marginnote{Levels} and map provision is executed on different hierarchical levels. \citet{Nothdurft2014} transferred the concept of \citet{Oberlander2008} to distinguish map information by topological, semantic, and metric properties to the field of automated driving. Based on \citet{Du2004}, \citet{Matthaei2015b} differentiated between macroscale (road-level), mesoscale (lane-level), and microscale (within lane) map information and localization in those maps.

On each level, localization\marginnote{Localization} sensors provide data input to obtain an absolute, global pose from localization algorithms. This information is often combined in Bayesian filtering approaches with inertial movement data (cf. blue data flow in Figure \ref{fig:Localization}) to provide a position even between position fixes from, e.g., a satellite-based localization sensor. Current approaches are differentiated by their depth of data fusion (loosely, tightly, ultra-tightly-coupled) and summarized by \citet{Skog2012} cited by \citet[p. 43]{Matthaei2015a1}.

An absolute global pose\marginnote{Retrieval} is used together with perceived environment features to obtain a map-relative pose estimation. This map-relative pose is used to \textit{retrieve} map information and to provide it to modules in the perception column in order to augment perceived information by map information. 

Depending\marginnote{Updating} on the implementation, a second data flow from the perception column towards the map provision column may exist. This is to use features and a concurrently obtained map-relative pose to update maps with perceived information. This concurrent map-relative localization and map-updating process may be repeated on the earlier introduced hierarchical levels. Information may be exchanged between the levels to keep maps consistent.

Different\marginnote{Limitations} technologies exist to serve the different (vertical) levels in the functional system architecture with different needs for accuracy. On a macroscale level (roads), global navigation satellite system solutions found in today's vehicle entertainment systems are largely sufficient. For mesoscale localization on a correct lane as well as for microscale localization within a lane, a higher accuracy is needed. Signal distortions in ionospheric layers can be compensated by utilizing different carrier frequencies and correction data from ground stations may be used to increase accuracy. Yet, accuracy as well as reliability are insufficient to serve as a single, non-redundant source for localization in automated driving. This becomes particularly obvious in urban environments or complex multi-level highway interchanges.

All\marginnote{Error Propagation} information from the localization and map-provision column is subject to errors in the localization as well as errors in the maps itself. At the time of writing there is no guarantee on information integrity and timeliness of data. Thus, incorrect localization or map data may possibly propagate to subsequent modules and compromise decisions and behavior. To ensure awareness of this, every module that uses map data is colored in violet. 

\subsubsection{Enhancements to the State of the Art}
The localization and map provision has been restructured. The \enquote{external data} and \enquote{absolute global localization} columns in \citet{Matthaei2015b} have been summarized into one \enquote{localization and map provision} column. \enquote{External data} was renamed to map provision to ensure that modules are activities, as in the Unified Modeling Language (UML) standard. Hence, \enquote{external data} is ---similar to, e.g., a \enquote{scene}--- a data container and thus an arrow rather than a module. We think that the level of abstraction for \enquote{localization} and \enquote{external data} seemed less aggregated than for example \enquote{perception} or \enquote{planning and control} which form other columns. Further, the titles of the map provision blocks have been changed towards what they actually provide: Maps. The \enquote{world modeling} used by \citet{Matthaei2015b} leaves room for confusing it with the term's connotation in the community to where it reflects activities which are here summarized under context modeling. \citet{Matthaei2015b} did not provide details within the map provision block. The refinements in \citet[p. 45]{Matthaei2015a1} within those blocks have now been incorporated to make them accessible to a non-German-speaking audience. 

\vspace{-2mm}
\section{Open Issues}\vspace{-2mm}
Despite long and intense discussions, there are still several open issues in the functional system architecture. Three aspects will be highlighted here: 

First of all,\marginnote{Context Modeling} the name of the context modeling seems counter-intuitive due to the fact that it only outputs a scene. Indeed, a scene is part of the context and according to its wide definition (cf. section \ref{sec:Background}), a full context may never be represented. Yet, certain context information may be used for better scene modeling. Thus, the term \enquote{context modeling} for the overall block seems more appropriate. 

Secondly,\marginnote{Additional Links} in architecture discussions with other research groups in the Uni-DAS society\footnote{Uni-DAS workshop on functional system architectures in October 2015 in Darmstadt, Germany. www.uni-das.de} the idea was voiced for feedback from stabilization modules towards model-based filtering modules. That is, to adopt models if, e.g., the ego vehicle is not following a planned trajectory when it is drifting.

Thirdly,\marginnote{Driver Monitoring} no clear answer has yet been provided where a driver or passenger\footnote{Might be necessary in a SAE level 5 system to help minors or elderly passengers for instance in case of a medical emergency situation or to ensure that they remain seated while driving.} monitoring camera should be located in the architecture. One could argue that it is irrelevant if an operator provides a maneuver input by a button or the camera and that it thus shall be part of the human machine interface. Likewise, it may be considered as a sensor and part of the perception column. A driver or passenger monitoring is so far not part of the Stadtpilot project.

Furthermore, an open point is the clear differentiation between \enquote{occupancy grid mapping} in the perception column and \enquote{feature map provision} in the localization and map provision column. Occupancy grid mapping is necessary in perception for local dynamic maps, free space extraction, or dynamic classification. If static elements are aggregated in a global feature map, it is part of the map provision column. Hence, the age of features to be typically still maintained in the grid or map is a distinguishing factor, but there is still room for a better distinction between both.    

In current discussions about the potentials and demands of automated vehicles, a server-based shared map is a key to the availability for automated vehicles. It is not explicitly modeled in the architecture, since we think it is part of the V2X connectivity. A more sophisticated integration into the architecture of the ego vehicle seems not helpful, as it would change the focus from the aspired architecture for a single automated vehicle towards an overall architecture for a whole traffic system. That would require several additional aspects like trusted authorities for information validation or traffic management authorities, which are out of scope of this article.

Moreover, there are still discussions on the point whether navigation or guidance has ultimate decision power if a planned route is followed or a route alternative is selected. If a traffic jam is detected, it is clearly a navigation task to adapt the route. Vice versa, if enforcing to take a highway exit would result in a collision, it is the tactical guidance layer that decides to not take the exit and to request a replanned route to reflect the reality of having missed that particular exit. There is a gray area in between where following the route is still within the specifications of what the automated vehicle \textit{can do}, but where in the given situation it is \textit{just now}, tactically a better choice to rather pick a route alternative with a minimal detour to avoid risk or maintain comfort goals. As in section \ref{sec:PlanningControl}, we see these decisions to be under the decision-making authority of the guidance level, but not without controversy.   

Another issue is where predictions are to be found in the architecture. To the authors, a prediction is rather a tool to be used in several modules. For instance, model-based filtering will use prediction models. Likewise, a situation prediction might be necessary in the guidance module or a movement prediction in the stabilization module. One could ask if there is a prediction even in the context model to provide not only the current but even future scenes. A possible way to illustrate predictions in the architecture could be to extend the two-dimensional architecture by a third dimension in which prediction is an additional layer. This comes to the price of visual distinctiveness and presentability. Another way could be to introduce multiple \textit{views} on the architecture for particular aspects.

Furthermore, the allocation of self representation to a particular block in the architecture is not as clear as it seems. For sure, it is mainly a bottom up process to aggregate information from vehicle sensors. Yet, execution monitoring might detect that a vehicle's deviation from its intended trajectory is high and thus the maneuver capabilities of that vehicle are limited. In other terms, there is goal and value specific information for self modeling in the planning and control column. Hence, certain aspects of self modeling could be spread over several hierarchical levels and columns in the architecture and thus limit the conceptual rigorousness that structure diagrams of the architecture suggest. Once more, a third dimension with a separate layer for self representation could alleviate this issue. In this layer not only the self representation, but also all forms of self monitoring and execution monitoring could be placed. The result could be aggregated in the scene/context model and used for decision making and control in the planning and control column.

Possibly not fully covered is the aspect of cooperation and competition between multiple agents. So far, implicitly cooperative behavior \cite{Ulbrich2013} and explicit Vehicle-To-Infrastructure communication \cite{Saust2012} has been implemented in the Stadtpilot project. Yet, it seems likely that future research on cooperation and competition may not be fully covered in the architecture. We assume an additional \textit{view} on the architecture might be required to cover these aspects with all its various facets.  

Last of all, the role of Vehicle-To-X communication is still subject to discussions. While the current communication column is eligible to broadcast information from the planning and control or perception column, an opposite communication flow for Vehicle-To-X data input is harder to incorporate. Currently, this induces a right-to-left information flow that contradicts the main signal flow direction otherwise going from left-to-right. A workaround would be once more to open a third dimension or additional \textit{view} for Vehicle-To-X communication as it has interfaces with many blocks. A possible implementation specific addition to the architecture could be a data flow from the decision modules to other traffic participants or the infrastructure via Vehicle-To-X and vice versa. E.g., the selected route, the selected maneuver as part of the situation for Vehicle-To-X communication, or a planned trajectory on the stabilization level.
 
\vspace{-2mm}
\section{Conclusions}
This article presented a refined functional system architecture for an automated vehicle. The concept of hierarchy and functional separation has been introduced and applied. The interfaces between the modules have been detailed and the modifications to the state of the art have been presented. 
To the authors, this functional system architecture is still an organic structure that will be modified and refined to address the open issues.
 
\appendices

\vspace{-3mm}
\section*{Acknowledgment}\vspace{-1mm}
The authors would like to thank the Stadtpilot team at TU Braunschweig for their valuable inputs and reviews of this article. We thank Volkswagen Group Research and the Auto-Pilot team for their financial and institutional support of the underlying research of this article. Further we thank the Federal Ministery for Economic Affairs and Energy for funding our research as part of the aFAS project, the German Research Foundation for supporting us in the Controlling Concurrent Change project, and the Daimler and Benz Foundation for funding the Value Based Decision Making project. We thank Prof. Mykel Kochenderfer for hosting Andreas Reschka at Stanford Intelligent Systems Lab while this article was written.
\vspace{-3mm}

\ifCLASSOPTIONcaptionsoff
  \newpage
\fi

\printbibliography

\begin{IEEEbiography}[{\includegraphics[width=1in,height=1.25in,clip,keepaspectratio]{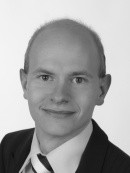}}]{Simon Ulbrich}
is currently working at Audi AG on piloted driving systems and is pursuing his PhD at TU Braunschweig at the Institute of Control Engineering. Before that, he finished a Diploma Degree in \textit{Electrical Engineering/Industrial Engineering} at TU Braunschweig and a Master of Science Degree in \textit{Industrial and Systems Engineering} at the Georgia Institute of Technology. His main research interests are tactical behavior planning and context modeling for automated driving.
\end{IEEEbiography}

\begin{IEEEbiography}[{\includegraphics[width=1in,height=1.25in,clip,keepaspectratio]{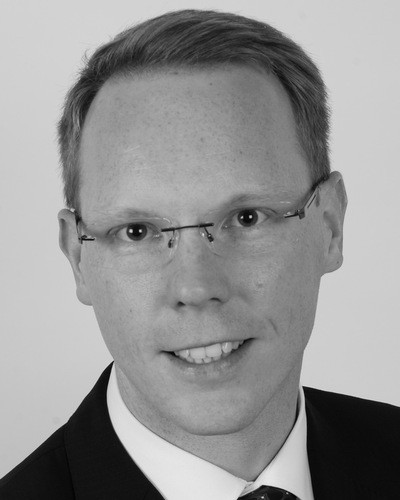}}]{Andreas Reschka}
is currently pursuing a PhD at the Institute of Control Engineering at TU Braunschweig and visiting the Stanford Intelligent Systems Laboratory. He holds a Bachelor of Science in \emph{Network Computing} from Technische Universität Bergakademie Freiberg and a Master of Science in \emph{Information Management and Information Technology} from Universität Hildesheim. His main research topics are self-awareness, safe behavior, functional safety, and development processes for autonomous vehicles.
\end{IEEEbiography}

\begin{IEEEbiography}[{\includegraphics[width=1in,height=1.25in,clip,keepaspectratio]{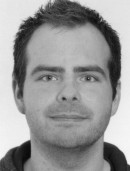}}]{Jens Rieken}
works as a research assistant at the Institute of Control Engineering at TU Braunschweig since 2012 and is currently pursuing his PhD. He holds a Master of Science Degree in \emph{Electrical Engineering} from TU Braunschweig. His main research topics are environment perception and scene understanding for urban scenarios, as well as designing algorithms for point cloud processing.
\end{IEEEbiography}

\begin{IEEEbiographynophoto}{Susanne Ernst}
is with the Institute of Control Engineering at TU Braunschweig since 2015 and is currently pursuing her PhD. She holds a Master of Science Degree in \emph{Mechanical Engineering} from TU Braunschweig. Her main research topics are behavior recognition of other traffic participants and decision making for automated vehicles.
\end{IEEEbiographynophoto}

\begin{IEEEbiography}[{\includegraphics[width=1in,height=1.25in,clip,keepaspectratio]{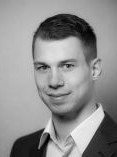}}]{Gerrit Bagschik}
works as a research assistant at the Institute of Control Engineering at TU Braunschweig since 2013 and is currently pursuing his PhD. He holds a Master of Science Degree in \emph{Computer and Communications Engineering} from TU Braunschweig. His main research topics are functional safety and scenario based hazard analysis for automated vehicles.
\end{IEEEbiography}

\begin{IEEEbiography}[{\includegraphics[width=1in,height=1.25in,clip,keepaspectratio]{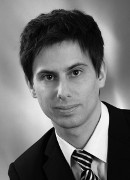}}]{Frank Dierkes} 
works as a research assistant at the Institute of Control Engineering at TU Braunschweig since 2013 and is currently pursuing his PhD. He holds a diploma degree in \emph{Computer Engineering} from RWTH Aachen University. His main research topic is behavior generation under uncertainty for automated vehicles.
\end{IEEEbiography}

\begin{IEEEbiography}[{\includegraphics[width=1in,height=1.25in,clip,keepaspectratio]{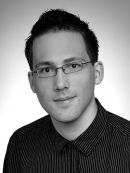}}]{Marcus Nolte}
works as a research assistant at the Institute of Control Engineering at TU Braunschweig since 2014 and is currently pursuing his PhD. He received his Master of Science in \emph{Electrical Engineering} from TU Braunschweig. His main research topics are self-awareness and motion planning for over-actuated automated vehicles.
\end{IEEEbiography}

\begin{IEEEbiography}[{\includegraphics[width=1in,height=1.25in,clip,keepaspectratio]{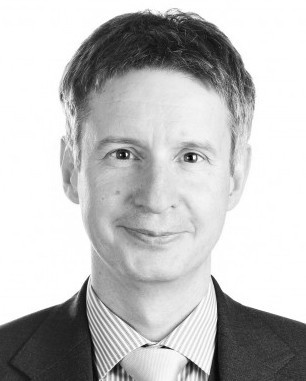}}]{Markus Maurer}
has held the chair for Vehicle Electronics at TU Braunschweig since 2008. His main research interests include autonomous road vehicles, driver
assistance systems and automotive systems engineering. From 2000 to 2007 he was active in the development of driver assistance systems at Audi AG.
\end{IEEEbiography}
\vfill
\end{document}